\documentclass[prb,twocolumn,preprintnumbers,superscriptaddress]{revtex4-2}
\usepackage{graphicx}
\usepackage{dcolumn}
\usepackage{bm}
\usepackage{verbatim}
\usepackage[usenames]{color}
\usepackage{stmaryrd}
\usepackage{float}
\usepackage{units}
\usepackage{textcomp}
\usepackage{amsmath}
\usepackage{commath}
\usepackage{amssymb}
\usepackage{esint}
\usepackage{ae,aecompl}
\usepackage[colorlinks=true,linkcolor=blue,citecolor=blue]{hyperref}
\usepackage{subfigure}

\begin{document}

\title{Enhanced Efficiency of Intermediate-Band Semiconductor Solar Cells \\ Embedded with Quantum Dot Superlattices}

\author{Naira Petrosyan}
\email{naira_petrosyan@ysu.am}
\affiliation{Department of Condensed Matter Physics, Yerevan State University, Alex Manoogian 1, 0025 Yerevan, Armenia}

\author{Lilit Yeganyan}
\email{lilit.yeganyan@ysu.am}
\affiliation{Department of Condensed Matter Physics, Yerevan State University, Alex Manoogian 1, 0025 Yerevan, Armenia}

\author{Aram Manaselyan}
\email{amanasel@ysu.am}
\affiliation{Department of Condensed Matter Physics, Yerevan State University, Alex Manoogian 1, 0025 Yerevan, Armenia}

\author{Vram Mughnetsyan}
\email{vram@ysu.am}
\affiliation{Department of Condensed Matter Physics, Yerevan State University, Alex Manoogian 1, 0025 Yerevan, Armenia}

\author{Vidar Gudmundsson}
\email{vidar@hi.is}
\affiliation{Science Institute, University of Iceland, Dunhaga 3, IS-107 Reykjavik, Iceland}

\author{Albert Kirakosyan}
\email{kirakosyan@ysu.am}
\affiliation{Department of Condensed Matter Physics, Yerevan State University, Alex Manoogian 1, 0025 Yerevan, Armenia}

\begin{abstract}

We present a multiscale approach for modeling an intermediate-band solar cell based on a GaAs/GaAlAs quantum dot superlattice of cubic symmetry. Our framework combines high-accuracy theoretical calculations of the superlattice band structure and miniband-related absorption coefficient with experimentally determined interband absorption data. The quantum-mechanically derived absorption spectrum is incorporated into a drift–diffusion transport model in COMSOL Multiphysics, where key processes, including thermal and radiative recombination, are taken into account. This integrated methodology enables realistic modeling of device performance. Our results identify an optimal superlattice constant of 14 nm, yielding a maximum solar cell efficiency of 13.3\%. Further increase in the superlattice constant enhances the miniband-related absorption peak but reduces the generation rate in the n-type region, resulting in a net efficiency decrease. The proposed approach, integrating theoretical, experimental, and computational components, provides a reliable framework for assessing solar cells based on quantum dot superlattices.

\end{abstract}

\maketitle

\section{INTRODUCTION}
The continuous growth of global energy consumption and the urgent need to reduce dependence on fossil fuels have made solar energy one of the most promising candidates for sustainable power generation. The fundamental principle of photovoltaic conversion relies on the unique property of semiconductors to absorb light and transfer part of the photon energy to charge carriers-electrons and holes. A semiconductor photodiode separates and collects these carriers, conducting the generated current in one direction \cite{Hylton_2013, Jabbar_2021}. Thus, a solar cell is essentially a semiconductor diode carefully designed to efficiently absorb sunlight and convert it into electrical energy \cite{Handbook, Rehman_2023, Lin_2023}.
Conventional solar cells typically include a front metallic grid that allows light to reach the semiconductor surface, an antireflective coating that enhances light transmission, and a $p-n$ junction that establishes the internal electric field required for carrier separation and extraction \cite{Handbook}.

Over the past decades, various types of solar cells have been developed, including crystalline and amorphous silicon cells, III--V compound semiconductor cells, chalcopyrite thin-film devices \cite{Intro_Solar_Cells_2016}, and dye- or quantum-dot-sensitized solar cells \cite{CdS_GaAs_2012}.
Despite remarkable technological progress, the conversion efficiency of a single-junction solar cell remains fundamentally limited by the loss of low-energy photons that are not absorbed and the thermalization of carriers excited by photons with energies well above the bandgap. These intrinsic losses have motivated the development of third-generation photovoltaic concepts, including tandem and multi-junction cells, hot-carrier devices, and quantum-dot (QD) based architectures that exploit nanostructured materials for advanced light management \cite{CdS_GaAs_2012,Klimov_2006,Antolin_CompRen_2012,CC_QD_IBSC_2023, Petrosyan}.

QDs are semiconductor nanostructures that confine charge carriers in all three spatial dimensions, resulting in discrete, atom-like energy levels \cite{Chakraborty}. Their optical and electronic properties can be precisely tuned by varying the dot shape and size, composition, and surface conditions \cite{Antolin_CompRen_2012,ACSPhotonics_2017_QDIB,Scaccabarozzi_PPV_2023}.
Due to quantum confinement, QDs exhibit enhanced optical absorption, potential for multiple exciton generation, and adjustable spectral response, making them attractive for next-generation photovoltaic applications \cite{Klimov_2006,QD_Dye_2010,ACSPhotonics_2017_QDIB,Scaccabarozzi_PPV_2023}.
Moreover, when QDs are arranged periodically within a host matrix, the tunneling of electrons and holes between QDs leads to the formation of minibands where extended electron states are possible. In this sense, periodically arranged QDs or quantum dot superlattices (QDSL) imitate optical properties of crystals but with significantly different transition energies \cite{Marti_TSF_2006,Aeberhard_OQE_2012,Tsai_TED_2017}.
Such superlattices provide controlled carrier transport and engineered optical transitions, enabling higher absorption efficiency and energy structures suitable for light energy conversion \cite{Aeberhard_OQE_2012}. Previous experimental and theoretical studies have demonstrated the feasibility of QD-based architectures with controllable properties in systems such as CdS/GaAs, AlGaInAs, and GaAs/AlGaAs, which exhibit strong optical absorption and efficient carrier extraction \cite{CdS_GaAs_2012,AlGaInAs_2012,Marti_TSF_2006,Linares_JAP_2011,Noda_JNOPM_2010, Mughnetsyan_2017, Aghchegala_2010}.

Embedding QDs in the optically active region of a semiconductor allows the introduction of additional energy states within the bandgap of the matrix material, forming an intermediate band. In the intermediate-band solar cell (IBSC) concept, sub-bandgap photons can contribute through a two-step pathway: electrons are first excited from the valence band to the intermediate band, and subsequently from the intermediate band to extended conduction-band states. This mechanism increases the photocurrent without compromising the open-circuit voltage, thereby enhancing the overall conversion efficiency \cite{Luque_PRL_1997,Marti_2001,DesignConstraints_2002,Luque_Marti_PIP_2001,PRB_2015_Kita}.
To implement this concept using QDs, the coupling between localized states must be engineered so that the resulting miniband acts as an intermediate set of states within the conduction band \cite{Marti_2001,Marti_TSF_2006,Aeberhard_OQE_2012}.
In addition, constraints on dot size, inter-dot spacing, and potential depth are crucial to ensure the separation of quasi-Fermi levels and to minimize nonradiative thermalization losses \cite{DesignConstraints_2002, Kirk_2013}.


In this work, we present a comprehensive theoretical and device-level model of an IBSC based on GaAs/Ga$_{0.7}$Al$_{0.3}$As QDSL of cubic symmetry, where each quantum dot supports a single confined electronic state. The miniband structure of the superlattice is obtained using the tight-binding approximation by incorporating tunneling interactions between neighboring quantum dots. The resulting electronic dispersion and corresponding wavefunctions are used to calculate the total optical absorption spectrum of the system taking into account transitions from the valence band to the conduction band and intermediate-band, allowing us to quantitatively evaluate their respective contributions to the overall photovoltaic efficiency. The quantum-mechanically derived absorption spectrum and the photogeneration rate are subsequently integrated into a drift-diffusion transport model implemented in Semiconductor Module of COMSOL Multiphysics to assess the device performance. The recombination and thermalization mechanisms are considered in the frame of the Shockley-Read Hall theory \cite{Vedel_2023}. 

By systematically varying the QDSL lattice constant $a$, we demonstrate that miniband engineering provides a powerful means of controlling the absorption edge and enhancing sub-bandgap photogeneration. Our results reveal a direct correlation between miniband width, optical transition strength, and photocurrent generation, leading to a significant improvement in the conversion efficiency compared to a Ga$_{0.7}$Al$_{0.3}$As p–n junction solar cell without QDSL. This multiscale framework establishes a direct physical link between quantum confinement and macroscopic photovoltaic performance, providing a solid foundation for the design and optimization of high-efficiency QDSL-based intermediate-band solar cells.
Our work complements the theoretical and computational studies on QDSL solar cells mentioned above. However, we believe that a precise theoretical calculation of the band structure, absorption spectrum, and generation rate, combined with the use of COMSOL Multiphysics to solve the drift–diffusion equations while accounting for the main recombination processes, provides reliable and realistic predictions for the I–V characteristics and overall efficiency of the solar cell.

The paper is organized as follows. Section II presents the theoretical framework and methodology. The results are discussed in Section III, and the main conclusions are summarized in Section IV.

\section{Theory}
\begin{figure}
    \includegraphics[width=0.4\textwidth]{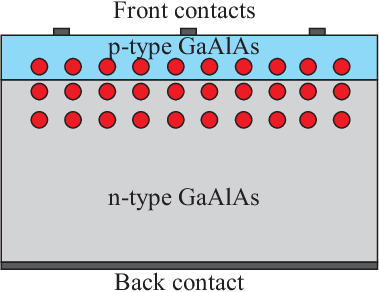}
    \vspace{-0.2cm}
    \caption{Schematic picture of the $p-n$ junction solar cell embedded with a QD superlattice.}
    \label{Structure}
\end{figure}

Consider a semiconductor $p-n$ junction solar cell constructed from layers of n-type and p-type Ga$_{0.7}$Al$_{0.3}$As with thicknesses $L_n$ and $L_p$, respectively, and embedded with a QDSL of cubic symmetry. The superlattice is composed of GaAs spherical quantum dots of radius $r_0$  (Fig.\ \ref{Structure}). The lattice constant of the QDSL is assumed, $a<<L_n,L_p$ so that the QDSL can be considered as infinite.

The QD superlattice embedded inside the solar cell creates an additional intermediate band inside the band gap of Ga$_{0.7}$Al$_{0.3}$As. Due to the cubic symmetry of the superlattice each elementary cell contains a spherically symmetric quantum dot with radius $r_0$. Each quantum dot has only a single bound state, due to its engineered size and material composition. The confinement potential of a single QD is modeled by a finite spherical square well:
\begin{equation}
U(x,y,z)=
\begin{cases}
-u, & 0< \sqrt{x^2+y^2+z^2} \le r_0,\\
0, & \sqrt{x^2+y^2+z^2}> r_0~,
\end{cases}
\label{eq:QDconf}
\end{equation}
where the energy is measured from the conduction-band edge of the barrier material Ga$_{0.7}$Al$_{0.3}$As and $u=0.6\Delta E_g=220$~meV is the conduction band offset of the GaAs/Ga$_{0.7}$Al$_{0.3}$As heterostructure.

The periodic potential of the  3D infinite superlattice is
\begin{equation}
    V(x,y,z)=\sum_{n_x,n_y,n_z=-\infty}^{\infty}{U(x-n_x a,y-n_y a,z-n_z a)},
    \label{eq:SLpotential}
\end{equation}
where $a$ is the lattice constant of the QDSL.

The periodic QD arrangement allows tunneling between neighboring sites, resulting in miniband formation. Within the tight-binding approximation, the miniband dispersion and electron wave functions are constructed from localized states of individual QDs. The localized electron wave function for the bound state of the spherical QD with potential (\ref{eq:QDconf}) is
\begin{equation}
\psi(r)=\frac{C}{\sqrt{r}}
\begin{cases}
J_{1/2}(k_1r), & 0< r \le r_0,\\
\dfrac{J_{1/2}(k_1r_0)}{K_{1/2}(k_2r_0)}K_{1/2}(k_2r), & r> r_0~,
\end{cases}
\label{eq:psi_total}
\end{equation}
where $C$ is the normalization constant, $J_l(x)$ and $K_l(x)$ are Bessel and modified Bessel functions, respectively, with $k_1=\sqrt{2m(\varepsilon_0+u)}/{\hbar}$ and $k_2=\sqrt{2m(-\varepsilon_0)}/{\hbar}$ and $m$ is the electron effective mass. The localized energy $\varepsilon_0$ can be determined from continuity conditions of the wave function (\ref{eq:psi_total}) and its derivative at $r=r_0$. The tight-binding miniband dispersion then reads \cite{Ziman_1972,Ibach_2009}
\begin{equation}
\varepsilon(\vec{k}) = \varepsilon_0 - B - 2A \left[ \cos(k_x a) + \cos(k_y a) + \cos(k_z a) \right],
\label{eq:dispersion_relation}
\end{equation}
with
\begin{displaymath}
    A=\!\!\int_\Omega\!\psi(x+a,y,z)\!\left(U-V\right)\!\psi(x,y,z)\,dxdydz, 
\end{displaymath}
\begin{displaymath}
     B=\!\!\int_\Omega\!|\psi(r)|^2\!\left(U-V\right)\,dxdydz.
\end{displaymath}
where the integration is carried out over the whole space, $\Omega$.
As a next step, we need to calculate the optical absorption spectrum of the system.  The experimental data of the absorption coefficient for the bulk Ga$_{0.7}$Al$_{0.3}$As alloy against the incident light wavelength corresponding to optical transitions from the valence band to the conduction band, can be found in \cite{Aspnes_1986}. The absorption coefficient of the QDSL corresponding to transitions from the valence band to the miniband, can be calculated using the formula:
\begin{equation}
  \alpha(\hbar\omega) = \alpha_0 \frac{1}{\hbar \omega} \frac{1}{N} \sum_{k_i,k_f} \left| \mathcal{M} \right|^2 \delta\!\left( E_f - E_i - \hbar \omega \right),
  \label{eq:abscoef}
\end{equation}
where
\begin{equation}
    \alpha_0=\frac{4\pi^2 e^2 \hbar}{n c m_0^2 \Omega_0}.
     \label{eq:alpha0}
\end{equation}
In Eqs.\ (\ref{eq:abscoef}) and (\ref{eq:alpha0}) $\Omega_0$ and $N$ are unit cell volume and number of unit cells, respectively, $\mathcal{M}$ is the transition matrix element, $m_0$ is the bare electron mass, $c$ is the speed of light and $n=\sqrt{\epsilon_r}$ is the refractive index.   

In the frame of dipole approximation the transition matrix element can be calculated using the expression:
\begin{equation}
    \left|\mathcal{M}\right|^2 =
    \frac{ \big\langle \left| P_{cv} \right|^2 \big\rangle \delta_{\vec{k}_i, \vec{k}_f}}{\Omega_0}
        \cdot \left| 
        \int_{\Omega_0}
    e^{i \vec{k}_i \cdot \vec{r}} \, \psi(\vec{r}) \, d\vec{r}
    \right|^2.
    \label{eq:MatrixElement}
\end{equation}
In Eq.\ (\ref{eq:MatrixElement}) $P_{cv}$ is the interband momentum matrix element calculated by means of the periodic Bloch functions for conduction and valence bands. Following \cite{I.Vurgaftman_2001}, in this work we choose $\big\langle \left| P_{cv} \right|^2 \big\rangle=m_0E_p/2$, where $E_p=28$eV is the Kane energy of the material. The presence of Kronecker delta symbol $\delta_{\vec{k}_i, \vec{k}_f}$ in Eq.\  (\ref{eq:MatrixElement}) demonstrates that only direct transitions are possible. During numerical calculations of absorption coefficient (\ref{eq:abscoef}) we assume parabolic dispersion for the valence band near $\Gamma$ point with heavy hole effective mass $m_{hh}$:
\begin{equation}
E_{i}(\vec{k}) = -E_{g} - \frac{\hbar^2 (k_x^2 + k_y^2 + k_z^2)}{2m_{hh}},
\label{eq:valence_dispersion_exact}
\end{equation}
and we used Eq.\ (\ref{eq:dispersion_relation}) for the dispersion of final state $E_f(\vec{k})$. Also Dirac delta function in Eq.\ (\ref{eq:abscoef}) is replaced by a Lorentzian of the form:
\begin{equation}
    L(\hbar \omega)=\frac{1}{\pi}\frac{\Gamma}{\Gamma^2+\left(E_f-E_i-\hbar \omega\right)^2},
    \label{Lorentz}
\end{equation}
where the parameter $\Gamma\approx26$meV describing the thermal widening of energy levels at room temperature $T=300$K.

 We use COMSOL Multiphysics software to calculate the photovoltaic characteristics of the system and to model the efficiency of the solar cells with and without embedded QDSL. The model of the system is constructed and the data of the solar spectrum and calculated total absorption spectrum of the system is imported into COMSOL. The photo-generation rate in the system is calculated using the expression \cite{Handbook}:
\begin{equation}
    G(x)=\frac{1}{h c} \int_{\lambda_{min}}^{\lambda_{max}}\lambda \alpha_{tot}(\lambda) F(\lambda)e^{-\alpha_{tot}(\lambda)x}d\lambda,
    \label{eq:Generation}
\end{equation}
where $\alpha_{tot}(\lambda)$ is the total absorption coefficient of the system and $F(\lambda)$ is the solar intensity spectrum.

In order to calculate electron and hole concentrations in the system, we solve the drift-diffusion equations numerically, taking also into account the direct and trap-assisted Shockley-Read-Hall recombination mechanisms \cite{Handbook}. Table I shows the values of all the parameters used in the numerical simulations.

\begin{table}
\caption{Numerical values of the parameters used in simulations.}
\begin{tabular}{ |l|c|c| } 
 \hline
 Electron effective mass & $m$ & $0.067m_0$ \\ 
 \hline
 Heavy hole effective mass & $m_{hh}$ & $0.45m_0$ \\ 
  \hline
  Radius of the dots & $r_0$ & $5$nm \\
 \hline
 Band gap for GaAs & $E_g^{GaAs}$ & 1420meV \\ 
 \hline
 Band gap for Ga$_{0.7}$Al$_{0.3}$As & $E_g$ & 1798meV \\ 
 \hline
 Length of the n-type GaAlAs & $L_n$ & $300\mu m$ \\
 \hline
 Length of the p-type GaAlAs & $L_p$ & $1\mu m$ \\
 \hline
 Cross-section area of the cell & $S$ & 1cm$^2$ \\
 \hline
 Donor concentration & $N_D$ & $10^{16}$cm$^{-3}$ \\
 \hline
 Acceptor concentration & $N_A$ & $10^{19}$cm$^{-3}$ \\
 \hline
 Electron and hole trap lifetime & $\tau_t$ & 5ns \\
 \hline
 Electron mobility & $\mu_n$ & 2300m²/(V·s) \\
 \hline
 Hole mobility & $\mu_p$ & 145.6m²/(V·s) \\
 \hline
 Relative permittivity & $\epsilon_r$ & 12 \\
 \hline
 Electron density of states  & $N_C$ & $6.53\times10^{17}$cm$^{-3}$ \\
 \hline
 Hole density of states  & $N_V$ & $1.12\times10^{19}$cm$^{-3}$ \\
 \hline
\end{tabular}
\end{table}

\section{Discussion of Results}

\begin{figure}
    \includegraphics[width=0.4\textwidth]{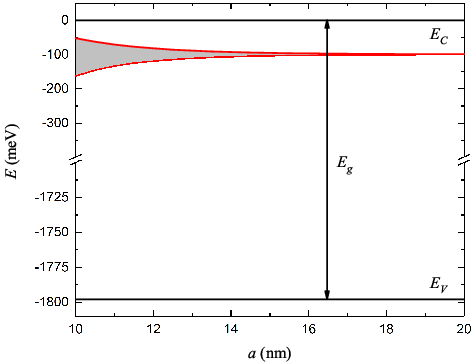}
    \vspace{-0.2cm}
    \caption{Band structure of the system versus lattice constant $a$.}
    \label{Miniband}
\end{figure}

Fig.\ \ref{Miniband} represents the energy band structure of considered system versus QDSL lattice constant $a$. For the values of $a>17$nm there is only a single energy level visible inside the bandgap of Ga$_{0.7}$Al$_{0.3}$As, which corresponds to the localized energy level of a single QD $\epsilon_0\simeq-99.5$meV. With the decrease of lattice constant $a$ this energy level transforms into a miniband due to quantum tunneling effects between the QDs. In the frame of tight binding approximation the energy dispersion of the miniband can be modeled by Eq.\  (\ref{eq:dispersion_relation}). The presence of the miniband inside the bandgap will change the optical absorption spectrum of the system and will allow the absorption of photons with smaller energy in solar spectrum.

\begin{figure}
    \includegraphics[width=0.4\textwidth]{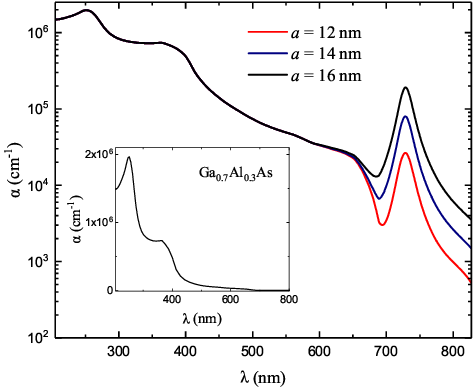}
    \vspace{-0.2cm}
    \caption{Absorption coefficient versus wavelength for various values of lattice constant $a$. Inset: Absorption coefficient of the bulk Ga$_{0.7}$Al$_{0.3}$As.}
    \label{Absorption}
\end{figure}

In Fig.\ \ref{Absorption}, the total absorption coefficients of the system are shown as functions of wavelength for three different values of the QDSL lattice constant, $a$. 
The curves correspond to transitions from the valence band to the conduction band and from the valence band to the miniband. 
Transitions from the miniband to the conduction band are not included, as their transition energies are too small and the corresponding wavelengths lie far beyond the solar spectrum. 
For comparison, the experimentally measured absorption coefficient of bulk Ga$_{0.7}$Al$_{0.3}$As (see~\cite{Aspnes_1986}) is shown in the inset of Fig.\ \ref{Absorption}. 
As evident from the plot, the presence of the miniband introduces additional absorption in the wavelength range of 650--800~nm. 
Moreover, with increasing lattice constant $a$, the absorption peak near $\lambda \approx 728$~nm associated with transitions from the valence band to the miniband becomes more pronounced.

\begin{figure}
    \includegraphics[width=0.4\textwidth]{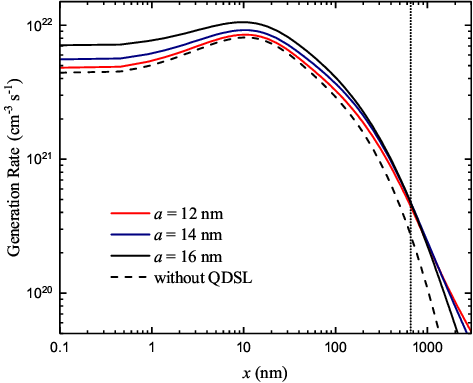}
    \vspace{-0.2cm}
    \caption{Generation rate versus penetration depth for various values of lattice constant $a$. Dashed line corresponds to the Ga$_{0.7}$Al$_{0.3}$As $p-n$ junction solar cell without QDSL.}
    \label{Generation}
\end{figure}

In Fig.\ \ref{Generation} the photogeneration rate of the system, obtained under standard illumination conditions at room temperature is presented against light penetration depth $x$ for three different values of the QDSL lattice constant, $a$. For comparison the photogeneration rate of the Ga$_{0.7}$Al$_{0.3}$As $p-n$ junction solar cell without QDSL is also presented as a dashed line. As one can see from the figure, the presence of QDSL inside the solar cell always increases the photogeneration rate for all values of $x$. For the small values of absorption depth the generation is higher for the case of $a=16$nm, due to stronger absorption in that case. With an increase of $x$, near the point $x\simeq660$nm, a crossing point between the generation curves can be observed for the cases $a=14$nm and $a=16$nm, because stronger absorption leads to a faster decrease of the light intensity with an increase of $x$. In order to describe the photovoltaic characteristics of our system we have to attach metallic Ohmic front contacts and back contact to the structure (see Fig.\ \ref{Structure}) and apply a voltage on them.  

\begin{figure}
    \includegraphics[width=0.4\textwidth]{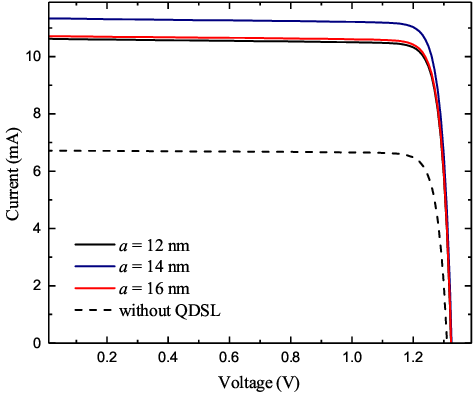}
    \vspace{-0.2cm}
    \caption{Volt--ampere characteristics for various values of lattice constant $a$. Dashed line corresponds to the Ga$_{0.7}$Al$_{0.3}$As $p-n$ junction solar cell without QDSL.}
    \label{VoltAmper}
\end{figure}

In Fig.\ \ref{VoltAmper} the Volt-Ampere characteristic curves of the system, obtained under standard illumination conditions at room temperature, are presented for three different values of the QDSL lattice constant, $a$. Again, for comparison the Volt-Ampere characteristic curve of the Ga$_{0.7}$Al$_{0.3}$As $p-n$ junction solar cell without QDSL is also presented as a dashed line. As one can see from the figure, the presence of QDSL inside the solar cell increases the short circuit current (the maximum current the cell can produce at $V=0$) substantially, but has very small impact on open circuit voltage (voltage when the circuit is open and current is 0). On the other hand, our simulations demonstrate that the highest value of short circuit current is obtained for the value of lattice constant $a=14$nm, which clearly indicates that this case could provide the highest efficiency of the solar cell. Interestingly, in the case of $a=16$nm the absorption coefficient related to valence band--miniband transitions is higher, but due to the weaker photogeneration rate in n-type region of the structure the generated photocurrent is lower.  

The solar cell efficiency is defined as the ratio of the maximum electrical power output to the incident light power input:
\begin{equation}
    \eta=\frac{P_{\textrm{max}}}{P_{\textrm{in}}},
\end{equation}
where $P_{\textrm{max}}=I_{\textrm{max}}\times V_{\textrm{max}}$ can be determined from the Volt-Ampere characteristic curve as the coordinates of the maximal power point, and the incident light power $P_{\textrm{in}}=1000\; \mathrm{W/m}^2\; \times S$ with $S$ being the area of the solar cell.

\begin{figure}
    \includegraphics[width=0.4\textwidth]{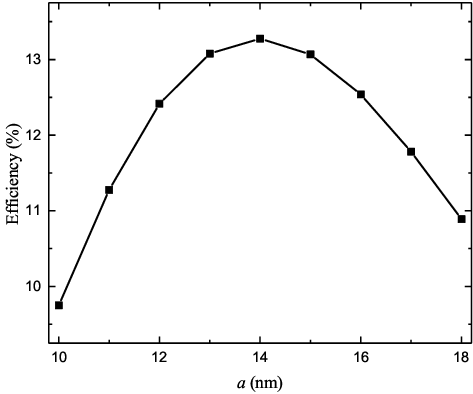}
    \vspace{-0.2cm}
    \caption{Solar cell efficiency versus lattice constant $a$.}
    \label{Efficiency}
\end{figure}

Following our COMSOL simulations, the efficiency of the Ga$_{0.7}$Al$_{0.3}$As $p-n$ junction solar cell without QDSL is $\eta\approx7.8\%$. In Fig.\ \ref{Efficiency} the efficiency of the Ga$_{0.7}$Al$_{0.3}$As $p-n$ junction solar cell embedded with QDSL is presented versus lattice constant $a$. As can be seen in the figure, the presence of the QDSL inside the solar cell can almost double its efficiency. As was already expected from the Volt-Ampere characteristics (see Fig.\ \ref{Generation}), for the considered model of the QDSL the maximum efficiency is obtained for the value of the lattice constant $a=14$nm and is equal to $13.3\%$.

\section{Conclusions}
In summary, we present a comprehensive multiscale approach for modeling an intermediate-band solar cell based on a quantum dot superlattice of cubic symmetry. 
Despite the fact that the bandgap of GaAs is better suited for solar radiation absorption, we consider Ga$_{0.7}$Al$_{0.3}$As bulk material with embedded GaAs quantum dots to avoid the effects of internal anisotropic strains on structural quality and to deal with an experimentally attainable size-, shape-, and periodicity-uniform distribution of quantum dots. 
Our approach combines a high accuracy tight-binding calculation of the superlattice band structure and the corresponding absorption coefficient related to the valence band to conduction miniband transitions. For the valence band to conduction band transitions we use experimentally determined data for the absorption coefficient.
The obtained absorption spectrum is subsequently incorporated into the drift-diffusion transport model implemented in Semiconductor Module of COMSOL Multiphysics, where all key processes, including thermal and radiative recombination, are taken into account. 
This integrated methodology enables realistic and predictive modeling of the device performance. 
Our results identify an optimal superlattice constant of 14 nm, yielding a maximum solar cell efficiency of 13.3\%. We show that incorporation of quantum dots into the bulk semiconductor leads of the efficiency and short circuit current enhancement of about 1.7 times. 
Interestingly, further increase in the superlattice constant enhances the miniband-related absorption peak but simultaneously reduces the generation rate in the n-type region (where the intensity of the penetrated light is significantly reduced), resulting in a net decrease in overall efficiency. The proposed approach, which integrates theoretical, experimental, and computational components, provides a reliable and physically consistent framework for assessing the performance of solar cells based on quantum dot superlattices.

\section*{Acknowledgement}

This research was supported by Higher Education and Science Committee of Armenia (Grants No. 24AA-1C004, 24LCG-1C004 and 21AG-1C048). The computations were performed at the Center for Modeling and Simulations of Nanostructures at Yerevan State University.

\section*{Author Contributions}

A.M. and V.M. supervised the project; N.P. and L.Y. performed the numerical calculations and simulations; all authors contributed to the analysis of the results and the preparation of the manuscript.


\end{document}